\documentclass[showpacs,prl,floatfix,reprint]{revtex4}
\usepackage{graphicx,amssymb,amsmath}
\begin{document}

\title{How congestion shapes cities: from mobility patterns to scaling}
%Congestion tears cities appart: a model for the polycentric transition}
%simple model for the monocentric-polycentric transition of cities}

\author{R\'emi Louf}
%\email{remi.louf@cea.fr}
\affiliation{Institut de Physique Th\'{e}orique, CEA, CNRS-URA 2306, F-91191, 
Gif-sur-Yvette, France\\
Email: remi.louf@cea.fr}

\author{Marc Barthelemy}
\affiliation{Institut de Physique Th\'{e}orique, CEA, CNRS-URA 2306, F-91191, 
Gif-sur-Yvette, France\\
Email: marc.barthelemy@cea.fr}

%% Abstract should be no more than 150 words for Sci Rep!
\begin{abstract}
The recent availability of data for cities has allowed scientists to exhibit scalings which present themselves in the form of a power-law dependence on population of various socio-economical and structural indicators. We propose here a stochastic theory of urban growth which accounts for some of the observed scalings and we confirm these predictions on US and OECD empirical data. In particular, we show that the dependence on population size of the total number of miles driven daily, the total length of the road network, the total traffic delay, the total consumption of gasoline, the quantity of $CO_2$ emitted and the relation between area and population of cities, are all governed by a single parameter which characterizes the sensitivity to congestion. Our results suggest that diseconomies associated with congestion scale superlinearly with population size, implying that --despite polycentrism-- cities whose transportation infrastructure rely heavily on traffic sensitive modes are unsustainable.
\end{abstract}

%\pacs{}

\maketitle

\section{Introduction}

The recent availability of an unprecedented amount of data has made possible quantitative studies of cities~\cite{Fujita:1999,Batty:2007,Marshall:2004}, opening the way to a Science of Cities. In particular, the discovery of allometric scaling relationships in cities has driven the quantitative research on urban systems in the past years. Indeed, there is a great amount of evidence to show that different socio-economic indicators in cities, such as the GDP, the crime rate, the number of patents as well as different structural indicators such as the total length of the road network, the urbanized land area, etc., exhibit robust scaling relationships with respect to population~\cite{Newman:1989,Makse:1995,Pumain:2006,Bettencourt:2007,Samaniego:2008,Rozenfeld:2008,Pan:2013}. The existence of these simple scaling relationship hints at the existence of universal processes shared by urban systems, and thus at the possibility of modeling cities.\\

A common trait shared by all complex systems --including cities-- is the existence of a large variety of processes occuring over a wide range of time and spatial scales. The main obstacle to the understanding of these systems therefore resides in uncovering the hierarchy of processes and in singling out the few ones which govern their dynamics. Albeit difficult, the hierarchisation of processes is of prime importance. A failure to do so leads to models which are  either too complex to give any real insight into the phenomenon, or too simple and abstract to have any resemblance with reality. As a matter of fact, despite numerous attempts~\cite{Fujita:1982,Makse:1995,Batty:2008,Frasco:2013,Bettencourt:2010,Bettencourt:2013}, a theoretical understanding of many observed empirical regularities in cities is still missing.\\

In the present study, we show that the spatial structure of the mobility pattern controls the behaviour of many quantities in urban systems. Indeed, cities are not only defined by the spatial organisation of places fulfilling different functions --shops, places of residence, workplaces, etc.-- but also by the way indivduals move among them. Understanding where people live, where and how they travel within the city thus appears as a necessary step towards a scientific theory of cities.\\ 

Although an increasing amount of data about mobility is now available~\cite{Gonzalez:2008}, we still lack a simple model explaining the dominant mechanisms governing the formation and evolution of mobility patterns. Many factors such as geographical constraints, facilities location and available transportation --to name a few-- can impact the mobility and it thus appears as an intricate issue. Here, we tackle the problem of mobility by making simplifying --yet not simple-- assumptions, trying to grasp the most important parameters which define the problem. We thus build upon a simple out-of-equilibrium model previously developed~\cite{Louf:2013}. This model, among other things, accounts for the polycentric transition of cities and gives a prediction for the number of centers as a function of population. We show that this framework allows us to predict the behaviour of many quantities related to mobility and the structure of cities: the scaling with population of the total time wasted in congestion, transport related $CO_2$ emissions, total travelled distance, total lane miles and surface area.

Our results allow us give a quantitative insight into two important debates around urban systems. First, we are able to discuss the benefits of polycentricity and quantify some of its aspects. Then, maybe more importantly, we are able to put into perspective the sustainability of urban systems.

\section{Results}

\subsection{Naive scalings}

We start by presenting some naive arguments to estimate the scaling exponents for the area $A$, the total daily distance driven $L_{tot}$ and the total lane miles $L_N$. Although these predictions turn out to be wrong, naive scalings are useful as a first approach to the problem as they allow us understand how the different quantities relate to one another.

\subsubsection{Surface area} 

First, we would like to estimate the dependence of the area $A$ of a city on its population $P$ --a long standing problem in the field~\cite{Makse:1995}. A first crude approach would be to assume that cities evolve in such a way that their population density $\rho = P/A$ remains constant. This assumption straighforwardly implies that the area should scale linearly with population
\begin{equation}
A \sim \lambda^2\, P
\label{eq:area_naive}
\end{equation}

where $\lambda^2$ is the average surface occupied by each individual (the assumption of a constant density is then equivalent to the one of a constant average surface per capita).

\subsubsection{Total length of roads} 

We would now like to estimate the total length $L_N$ of all the roads within a city. If we consider that the network formed by streets is such that all the nodes (intersections) are connected to their closest neighbour, the typical length of a road segment is given by

\begin{equation}
\ell_R \sim \sqrt{\frac{A}{N}}
\end{equation}

where $N$ is the number of intersections. Previous studies of road networks in different regions, and over extended time periods~\cite{Strano:2012,Barthelemy:2013}, have shown that the number of intersections is proportional to the population size. Therefore, the typical length of a road segment (between two intersections) varies with the population size $P$ as
\begin{equation}
\ell_R \sim \sqrt{\frac{A}{P}}
\end{equation}
and the total length of the network $L_N \sim P\ell_R$ should then scale as
\begin{equation}
\frac{L_N}{\sqrt{A}}\sim\sqrt{P}
\end{equation}
Using the naive scaling for the dependence of $A$ on population size given previously in Eq.~\ref{eq:area_naive} we finally get
\begin{equation}
L_N \sim P
\end{equation}

\subsubsection{Total daily commuting distance} 

\paragraph{Individual constraint} We would also like to estimate the total commuting distance $L_{tot}$. The first constraint on this distance comes from individuals's limitations and behaviour. We make here the simple assumption that individuals choose their residence and work place such that their total commuting distance is fixed (or at least smaller than a certain value) and equal on average to $\ell_C$. In that case, we would simply have
\begin{equation}
 \frac{L_{tot}}{P} \sim \text{const.}=\ell_c
\label{eq:assum}
\end{equation}
(by constant, we mean independent from the population size of the city).

\paragraph{The city structure constraint} An additional contraint on $L_{tot}$ is given by the structure of the city~\cite{Samaniego:2008}. Indeed, the individual commuting distance is also related to the total suface area of the city and the location of activity centers.

If we first assume that the city is monocentric, individuals are all commuting to the same center and the typical commuting distance $\ell^m_c$ is controlled by the typical size of the city of order $\sqrt{A}$ 
\begin{equation}
\frac{L_{tot}^{m}}{\sqrt{A}} \sim P
\end{equation}

On the other hand, if we assume that the city is completely decentralized, the typical commuting distance is of order the nearest neighbour distance $\sqrt{A}/\sqrt{P}$, and we obtain
\begin{equation}
\frac{L_{tot}^{d}}{\sqrt{A}} \sim \sqrt{P}
\end{equation}

\subsection{Comparison of naive scalings with empirical results}

The comparison of the naive exponents with the exponents measured on US data is shown in Table 1 (see the Methods section for details about the data). There are important discrepancies, which we discuss in the following.

First, we note that the naive scaling for the surface area $A$ predicts a value of the exponent that is quantitatively --and worse, qualitatively-- different from that observed. Indeed, we find that for real cities
\begin{equation}
A \sim P^{\,a}
\end{equation}
with $a=0.85$. While the naive argument implies a linear dependence of the surface area $A$ with population, we find a sublinear scaling in the data, which is a qualitatively different behavior (Table~\ref{table:naive}). This disagreement on this basic quantity will naturally impact the scaling of the other quantities.

%On the other hand, the prediction for $L_N / \sqrt{A}$ seems better, with an exposant close to the one observed. This %implies that the link between the network structure and the population is well captured by the naive argument. However, %the naive prediction for $L_N$ is incorrect: it predicts a linear dependence on population while data indicate a sublinear %behaviour. This confirms the problem with the naive prediction for the behaviour of $A$ that we discussed above.

\begin{figure}
\includegraphics[width=1.0\textwidth]{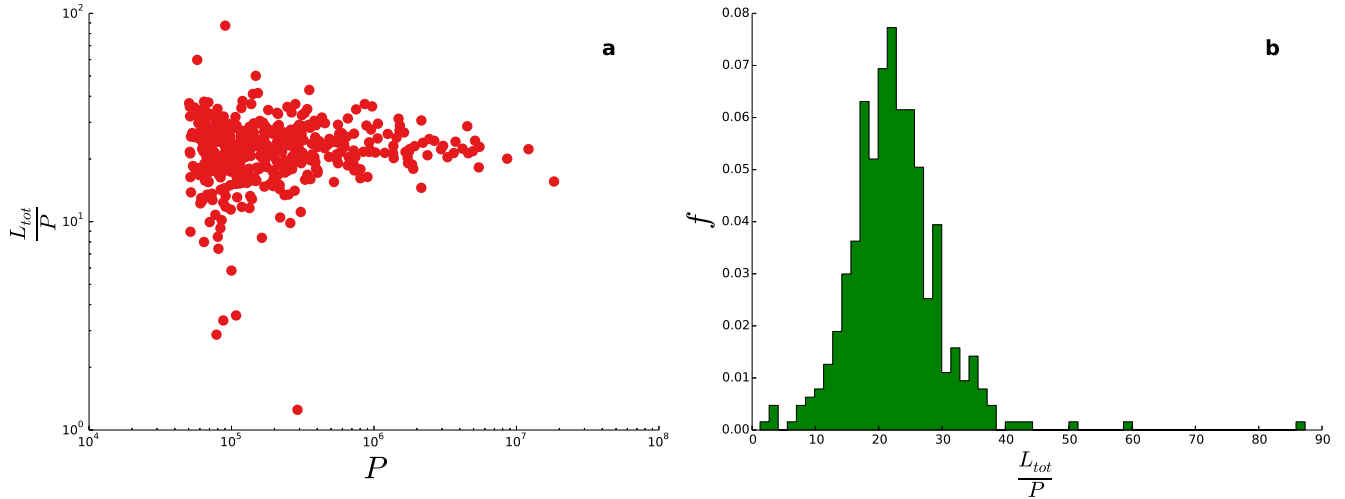}
\caption{Constant daily driven distance per capita. (a) daily total driven distance per capita as a function of population for 441 urbanised area in the US in 2010. The data shown in the plot are compatible with a population-independent behaviour. (b) Histogram of the daily total driven distance per capita for the same cities. The average daily driven distance for these cities is $23$ miles, and the standard deviation $7$ miles.}
\label{fig:LtotoverP}
\end{figure}

The data also show that $L_{tot}/P$ can be considered reasonably independent from $P$ (with a value of approximately $23$ miles for the US, see Fig.~\ref{fig:LtotoverP}), in agreement with the individual constraint assumption (Eq. \ref{eq:assum}). This finding is also in agreement with the results drawn from census data in Germany by~\cite{Wilkerson:2013}. Although this assumption of a constant distance is simple and verified on the US data, we think that it deserves to be systematically tested on other datasets for other countries and cities. 

Finally, the scaling of $L_{tot}/\sqrt{A}$ given in the extreme cases of a monocentric city structure and a totally decentralized city structure disagree with the value measured on data (see Table~\ref{table:naive}). This suggests that most cities have a structure that is neither completely centralized, nor totally decentralized. In particular, this result cast some doubts about the study \cite{Bettencourt:2013} which assumes implicitely that cities are always monocentric.
Any situation between the two previous extreme cases would give a scaling of the form $L_{tot}/\sqrt{A}\sim P^{\,b}$ where $b \in [1/2,1]$. One can easily see that this expression is consistent with that of $A/\lambda^2$ and $L_{tot}/P$ if
\begin{equation}
b = 1-\frac{a}{2}
\label{eq:consis}
\end{equation}
which is indeed what we observe empirically (up to error bars). This preliminary analysis thus leads us to the conclusion that, in order to compute the various exponents, we need to better describe the structure of commuting patterns. In other words, we need to find a description of cities that goes beyond the naive monocentric or totally decentralized views, and which accounts for the observed sub-linear scaling of the surface area $A$.

% The choice of assumptions is crucial here and is essentially determined by the period considered, and the time scales associated with the evolution of the urban system. Mobility indeed depends on transportation means which evolved very rapidly during the last two centuries. Because the area of an urban system results from centuries of evolution, we do not a priori expect our model --where individual vehicles are assumed to be the only vector of mobility-- to give a prediction valid for all countries and all times. At the present stage, a full discussion about this problem and even the possibility of describing the time evolution of cities might seem within reach, but we leave it for further studies. 
\subsection{Beyond naive scaling: modeling mobility patterns}

We begin with the assumption that mobility patterns are mostly driven by the daily commuting and we would like to understand how an individual, given his household location,  will choose his job location. We assume that this choice will be determined by two dominant factors: the expected wage at a given job, and the commuting time to this job's location. Indeed, places with high average salaries are attractive, but having to spend a sensible amount of time commuting every day is less desirable. We assume there are $N_c$ potential activity centers in the city, each characterized by an average wage $w(j)$ at location $j$. This wage is endogenously determined and depends a priori on many factors such as agglomeration effects~\cite{Glaeser:2001}, the type of industry, etc. Although it is in principle possible to write down equations to determine the wage (as attempted in~\cite{Fujita:1982} for instance), not only is it impossible to solve them, but also not necessarily useful. A similar situation arises in physics when one studies the behaviour of atoms made of a large number of electrons. Physicists found out~\cite{Dyson:1962} that, in fact, a statistical description of these systems relying on random matrices could lead to predictions which agree with experimental results. We would like to import this idea of replacing a complex quantity such as wages --which depends on so many factors and interactions-- by a random one in spatial economics. So, we treat the wage \emph{as if} it was exogenous and random~\cite{Louf:2013}, that is we write $w(j)=s\,\eta_j$ where $s$ represents the typical income in this city and $\eta$ is a random number chosen uniformly in $[0,1]$. Furthermore, we assume that the commuting time does not only depend on the distance between the two places, but also on the traffic $T_{ij}$ between those two locations. An individual living at $i$ will thus commute to the center $j$ which corresponds to the best trade-off between income and commuting time, thus to the center $j$ such that the quantity
\begin{equation}
Z_{ij} = \eta_j - \frac{d_{ij}}{\ell} \left[ 1 + \left( \frac{T(j)}{c} \right)^\mu \right]
\end{equation}
is maximum~\cite{Louf:2013}. The quantity $d_{ij}$ is the euclidean distance between $i$ and $j$ (both supposed to be scattered randomly across the city), $T(j)$ the total incoming traffic at $j$, $c$ the capacity of the underlying transportation network, and $\mu$ is an exponent describing the sensitivity of the network to congestion. The quantity $\ell$ is the maximum distance that people can financially travel daily, defined as the ratio between the typical individual income and the transportation costs per unit of distance.\\

This simple model displays a surprisingly rich behaviour~\cite{Louf:2013}. In particular, it accounts for the monocentric to polycentric transition observed in most cities. It has been a well-known fact for quite some time that as cities grow, they evolve from a monocentric organisation where all the activities are concentrated in the same geographical area --usually the central business district-- to a more distributed, polycentric organisation~\cite{Fujita:1982,McMillen:2003}. Several theories in spatial economics exist~\cite{Fujita:1999}, but are not satisfactory for many reasons. Among other things, they do not take congestion into account and have no predictive, testable content~\cite{Bouchaud:2008}. Within this framework, congestion is actually responsible for the transition, and the number of activity centers in a city of population size $P$ is on average given by
\begin{equation}
k = \left( \frac{P}{P^*} \right)^{\frac{\mu}{\mu+1}}
\label{eq:num_centers}
\end{equation}  
with 
\begin{equation}
P^* = c \left( \frac{\ell}{\sqrt{A} N_c} \right)^{1/\mu}
\end{equation}

% Forgetting all the other variables but $A$ and $P$ we get

% \begin{equation}
% k \sim \left( P\,A^{1/2\mu}\right)^{\frac{\mu}{\mu+1}}
% \label{eq:num_centers}
% \end{equation}  

Using data of employment per Zip Code Area in the US~\cite{Louf:2013}, we showed that

\begin{equation}
k \sim P^{\,\alpha}
\end{equation}

where we measure $\alpha = 0.64 \pm 0.12$ ($95\%$ confidence interval (CI)). In other words, the number of centers scales sublinearly with population size.

% We now show how the polycentric structure that emerges from the existence of congestion controls the dependence of the surface area on population. 
%
\begin{figure*}
\includegraphics[width=1.0\textwidth]{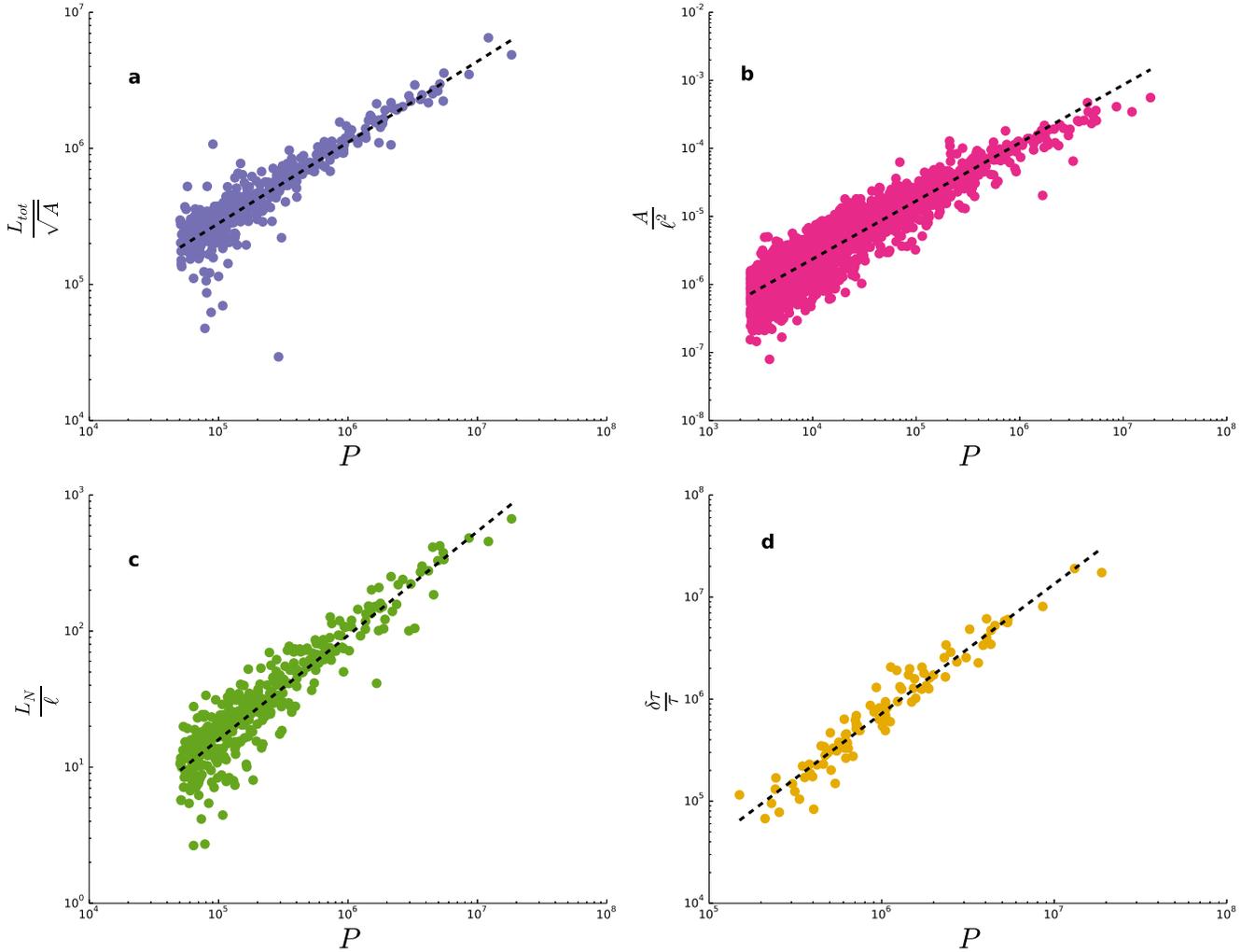}
\caption{Mobility and city structure and their impact on agglomeration economies and diseconomies. (a) Variation of the daily total driven distance with the population for 441 urbanized areas in the US in 2010. The dashed line shows the power-law fit with exponent $0.595 \pm 0.026$ ($r^2=0.90$). (b) Variation of the land area with population for 3540 urbanised areas in the US in 2010. The fit assuming a power-law dependence gives an exponent $0.853 \pm 0.11$ ($r^2=0.93$). Both exponents are smaller than 1, as predicted by our theory. (c) Variation of the total lane miles with population for 363 urbanised areas in the US. A power law fit  (dashed line) gives $L_N / \ell = P^{\,0.765 \pm 0.033}$ ($r^2=0.92$). The sublinear behaviour --which agrees with our prediction-- means that larger cities need to spend less in infrastructure per capita than smaller ones. (d) Variation of the total delay due to congestion with population for 97 urbanised areas in the US. A power law fit gives an exponent $1.270 \pm 0.067$ ($r^2=0.97$). The superlinear behaviour agrees with the prediction given by our model and challenges the claims of sustainability of cities.}
\end{figure*}

\subsection{Computing the exponents}

\subsubsection{Area}

At this stage, the number of centers is a function of population and the area

\begin{equation}
k = F\left(A,P\right)
\end{equation}

and we need an additional equation in order to get a closed system. Here we focus on the area and its evolution with the population size, which reflects the growth process of the city. In the following, we will investigate two different approaches. It is worth noting that both approaches give results in qualitative agreement, showing that some stylized facts ---such as super- or sublinearity--- are very robust.\\ 

\paragraph{Fitting procedure.}

In the absence of knowledge of the processes responsible for urban sprawl, we can assume that the area behaves as 
\begin{equation}
A \sim P^{\,a}
\label{eq:fit}
\end{equation}
where $a$ is the exponent to be determined, through fits on data. The empirical value for the exponent for the US data is $a\simeq 0.85$. Once this exponent is given we can then compute the various exponent for the quantities of interest (see the following and table 2). We get for the number of centers $k$

\begin{equation}
k \sim P^{\frac{\mu + a/2}{\mu+1}}
\end{equation}

which is sublinear as long as $a<2$, in agreement with the empirical results for US cities. As we will see, this approach yields the same qualitative behaviours as those predicted with the method of the next section. In other words, even if the main mechanism behind urban sprawl is not congestion, the conclusions of this paper are not affected as long as the area scales \emph{sublinearly} with population.\\

\paragraph{Coherent growth.}

Let us now assume that the scaling of $A$ with population is determined by the number of activity centers and the constant commuting length of individuals. This means that the growth of the area is controlled by the appearance of new activity centers. if we assume that a city is organized around $k$ activity centers and that the attraction basin of each of these centers are spatially separated~\cite{Louf:2013}, we then have  $A \sim k\, A_1$ where $A_1$ is the area of each subcenter's attraction basin. This area $A_1$ is related to the average individual commuting distance by $\sqrt{A_1} \sim L_{tot} / P$, and we obtain
\begin{equation}
A \sim k\,  \left( \frac{L_{tot}}{P} \right)^2 = k\, \ell_c^2
\label{eq:area_poly}
\end{equation}
This leads to expression for the number of centers
\begin{equation}
k \sim P^{\frac{2 \mu}{2\mu+1}}
\end{equation}

which is always smaller than $1$, also in agreement with the empirical results for US cities. We can now also compute the scaling of the surface area
\begin{equation}
\frac{A}{\ell_c^2} \sim \left( \frac{P}{c} \right)^{\frac{2 \mu}{2\mu+1}}
\end{equation}
We further assume that $L_{tot} / P$ is a fraction of the longest possible journey $\ell$ individuals can afford, that is to say 
\begin{equation}
\ell_c \sim \ell
\end{equation}
It is important to note that if $\ell_c$ is independent from $\ell$, the quantitative predictions of our model would still hold. 
The final expression for the area is then here given by
\begin{equation}
\frac{A}{\ell^2} \sim \left( \frac{P}{c} \right)^{\,2\,\delta}
\label{eq:area}
\end{equation}
where $\delta=\frac{\mu}{2\mu+1}$. The exponent $\delta$ is smaller than $1/2$ whatever $\mu\geq 0$, which implies that the density of cities increases \emph{sublinearly} with population. In other words, the density of cities \emph{increases}  with population. We verify this prediction in Table 2, with data about land area of urbanized areas in the US (Figure 2). We find $2 \delta_{emp} = 0.85 \pm 0.01\; (95\%\, CI)$ which is not too far from the theoretical value $2\delta_{th} = 0.64 \pm 0.12\; (95\%\, CI)$, equal to $\alpha$ in this case.

Because the area of an urban system results from centuries of evolution, we do not a priori expect our model --where individual vehicles are assumed to be the only vector of mobility-- to give a prediction valid for all countries and all times. Nevertheless, these results give us reasons to believe that the spatial structure of the journey-to-work commuting should still be the dominant factor in the dependence of land area on population.

\subsubsection{Total commuting distance}
%Ltot
Using Eq.~\ref{eq:assum} and Eq.~\ref{eq:area} we are now able to compute $L_{tot}/\sqrt{A}$
\begin{equation}
\frac{L_{tot}}{\sqrt{A}} = P\; \left(\frac{P}{c}\right)^{-\delta}
\label{eq:travelled_length}
\end{equation}
We plot $L_{tot} / \sqrt{A}$ for urbanized areas in the US on Figure 2, and one can check in Table 2 that the exponent predicted from the previously measured value of $\alpha$ agrees well with the exponent measured on the data. In the fitting case, the exponent would simply be given by $1-a/2$.

\subsubsection{Total length of roads}
%LN
If we use the previously derived expression for the area $A$, we find
\begin{equation}
L_N \sim \ell \; \sqrt{P}\; \left(\frac{P}{c}\right)^{\,\delta}
\end{equation}
The quantity $\delta$ is less than $1/2$, which implies that $L_N$ scales \emph{sublinearly} with the city's population size. In other words, larger cities need less roads per capita than smaller ones: we recover the fact that agglomeration of people in urban centers involves economies of scale for infrastructures. Within the fitting assumption (Eq.\ref{eq:fit}), we would obtain $(1+a)/2$.

\subsubsection{Total delay due to congestion}

Unfortunately, agglomeration in cities does not only generate economies. Congestion, for instance, is a major diseconomy associated with the concentration of people in a given area. A simple way to quantify the impairement caused by traffic congestion is through the total delay it generates. If we make the first order approximation that the average free-flow speed $v$ is the same for everyone, the total delay due to congestion is given --according to our model-- by
\begin{equation}
\delta \tau = \frac{1}{v} \sum_{i,j} d_{ij} \left( \frac{T_j}{c} \right)^\mu
\end{equation}
If we assume that all the centers share the same number of commuters --a reasonable assumption within our model~\cite{Louf:2013}-- we obtain
\begin{equation}
\delta \tau \sim \frac{L_{tot}}{v} \left( \frac{P}{k} \right)^{\mu}
\end{equation}
which, using the expressions for $L_{tot}$ and $A$ given in Eq.~\ref{eq:travelled_length} and Eq.~\ref{eq:area} respectively, gives
\begin{equation}
\delta \tau \sim \frac{\ell\; P}{v}\; \left(\frac{P}{c}\right)^{\delta}
\end{equation}

The total commuting time corresponding to the same distance but without congestion scales as $\tau_0\sim L_{tot}$ and thus less rapidly than the total delay which scales \emph{super-linearly} with population (even when polycentricity is taken into account). This means that, for the largest cities, delays due to congestion actually dominate the time spent in traffic, and that economical losses \emph{per capita} due to the time lost in congestion --and the corresponding strain on people's life-- increase with the size of the city. 

In the fitting assumption Eq. \ref{eq:fit}, and using the same arguments for the calculation of $\delta\tau$, we easily obtain for the exponent the value $1+\frac{\mu}{\mu+1}\left(1-\frac{a}{2}\right)$.

\subsubsection{Transport related $CO_2$ emission. Gasoline consumption}

Another diseconomy associated with congestion is the quantity of $CO_2$ emitted by cars and the gasoline consumed by motor vehicles. This amount not only depends on the distance that has been driven, but also on the traffic during the journey. It indeed turns out that for the same length driven, a car burns more oil when the traffic is heavy than when the road is clear.  Within our model, the presence of traffic is seen in the time spent to cover a given distance, and we write that the quantity of $CO_2$ emitted by a vehicle is proportional to the total time spent in traffic, leading to
\begin{equation}
Q_{CO_2}  = q \sum_{i,j} d_{ij} \left[ 1+ \left( \frac{T_j}{c} \right)^\mu \right]
\end{equation}
where $q$ is the average quantity of $CO_2$ produced per unit time. In the polycentric case with $k=k(P)$ subcenters, the typical trip 
length $\overline{d_{ij}}$ is given by $\sqrt{A/k}$ and we obtain
\begin{equation}
Q_{CO_2} = q\, \ell\, P \left[ 1 + \left(\frac{P}{c}\right)^{\delta} \right]
\end{equation}
The first term in brackets is a constant, and the quantity of $CO_2$ is thus dominated by congestion effects at large populations
\begin{equation}
Q_{CO_2} \sim q\; \ell\; P \left(\frac{P}{c}\right)^{\delta}
\end{equation}
and the total daily transport-related $CO_2$ emission per capita thus scales as 
\begin{equation}
\frac{Q_{CO_2}}{P} \propto  q\ell \left(\frac{P}{c}\right)^{\delta}
\end{equation}
The quantity of $CO_2$ emitted per capita in cities thus increases with the size of the city, a consequence of congestion. This prediction agrees with the exponent we measure (Figure 3) on data gathered for US and OECD cities (Data about the area and population of urbanised areas can be found  on the Census Bureau website \cite{DataUSA1}, data about congestions in urban areas can be found in the Urban Mobility Report \cite{DataUSA2}, and data about the total lane miles and the daily total miles driven in urbanized areas can be found on on the Federal Highway administration website \cite{DataUSA3}). We are aware that the scaling of $CO_2$ with population size is controversial, with results varying from one study to another. Although a systematic meta-analysis of these results is beyond the scope of this paper, we note that the authors of~\cite{Fragkias:2013} are concerned with the total emissions of $CO_2$, while this paper is only concerned with emissions due to transportations. Moreover, our prediction agrees well with the exponent of $1.33$ measured by the authors of~\cite{Oliveira:2014} on the same dataset, but with a different definition of cities. Finally, our prediction also agrees with measurements made in \cite{Rybski} for developing countries. 
\begin{figure}
\includegraphics[width=0.9\linewidth]{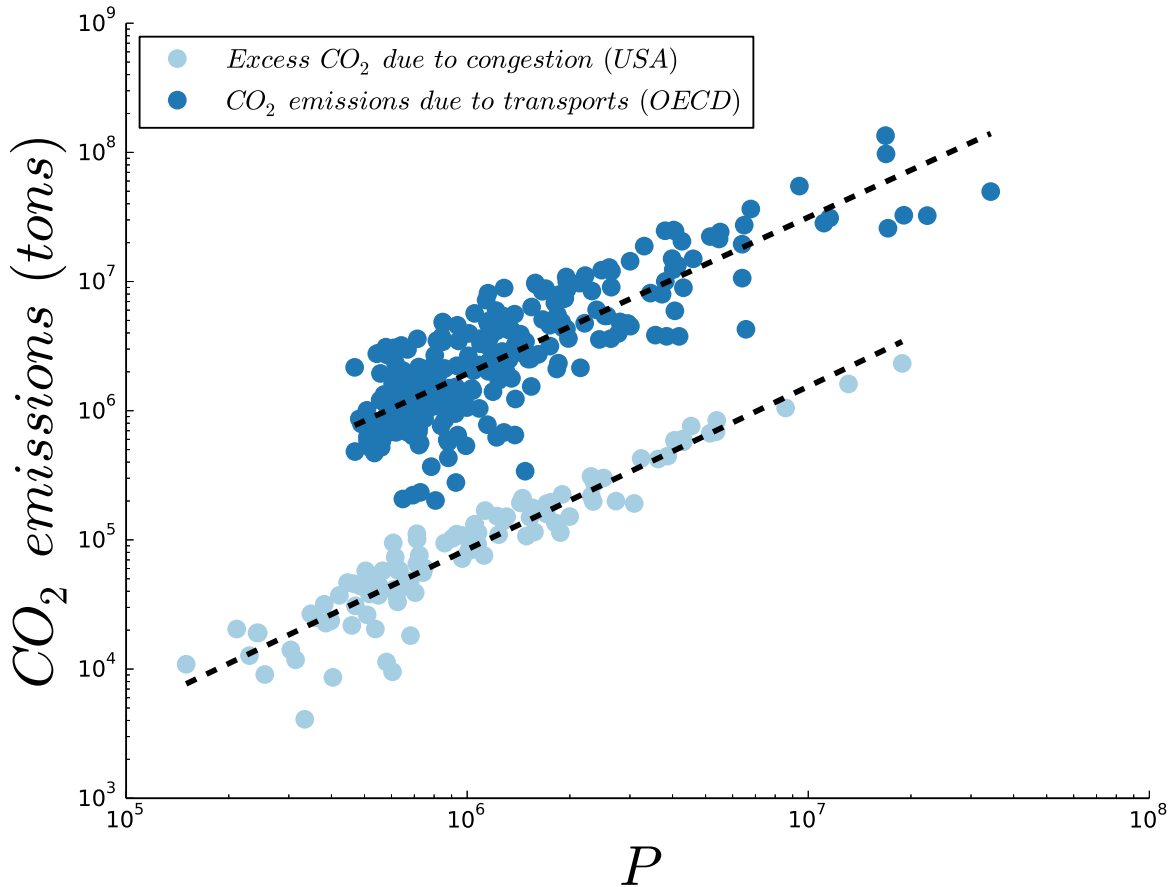}
\caption{Variation of $CO_2$ emissions due to transport with city size. In blue, excess $CO_2$ (in tons) due to congestion, as given by the Urban Mobility Report (2010) for 101 metropolitan areas in the US~\cite{DataUSA2}. In green, we show the estimated $CO_2$ emissions (in tons) due to transports, as given by the OECD for 268 metropolitan areas in 28 different countries (Data about the total $CO_2$ emissions due to transportation in major metropolitan area in the OECD can be found online~\cite{OECD}). The dashed yellow lines represent the least-square fit assuming a power-law dependency with multiplicative noise, which gives respectively $Q_{CO_2} \sim P^{1.262 \pm 0.089} (r^2=0.94)$ for the US data and $Q_{CO_2} \sim P^{1.212 \pm 0.098} (r^2=0.83)$ for the OECD data.}
\end{figure}

Another important related quantity is the the consumption of gasoline which in principle is proportional to the emission of $CO_2$ and the time spent driving. The total daily gasoline consumption is then given by
\begin{equation}
Q_{gas} \sim q\; \ell\; P \left(\frac{P}{c}\right)^{\delta}
\end{equation}
where $q$ is the average quantity of gasoline needed per unit time. From this expression, we see that the total daily gasoline consumption per capita scales as
\begin{equation}
\frac{Q_{gas}}{P}\sim \ell\, \sqrt{\frac{P}{\rho}} = \ell \sqrt{A}
\end{equation}
and is therefore not a simple function of the city density, in contrast with what was suggested by the seminal paper of Newman and Kenworthy~\cite{Newman:1989}. At this stage however, more data about gasoline consumption is needed to test this prediction and draw definitive conclusions.

\section{Discussion}

\subsection{Monocentric versus polycentric}

Although polycentricity emerges naturally from our model as a result of congestion, many circumstances can prevent or foster the appearance of new activity centers in a city. There are many debates as to whether policies should favour polycentric or monocentric developpement of cities. Most of them are based on ideologies and opinions about how cities should be, very few are based on a quantitative understanding of the city as a complex system. Although this only represents a small part of the debate, our model allows to quantify the effect of polycentricity on the total delay due to congestion.

We can indeed compute the total delay due to congestion in the case of a monocentric configuration. In this situation, all the population commutes to a single destination $1$ and we have
\begin{equation}
\delta \tau_{mono} = \frac{1}{v}\; \sum_i d_{i1} \left(\frac{P}{c} \right)^\mu = L_{tot} \left(\frac{P}{c} \right)^\mu
\end{equation}
It follows, using the expression given above for $L_{tot}$
\begin{equation}
\delta\tau_{mono} = \frac{\ell}{v}\; P^{1+\mu}
\end{equation}
From the fact that $1+\mu > 1+\frac{\mu}{2\mu+1}$, we indeed find that the total delay due to
congestion is worse for monocentric cities than it is for polycentric cities with the same population, which agrees with the usual intuition. More precisely the ratio of delays is given by
\begin{equation}
\frac{\delta\tau_{mono}}{\delta\tau_{poly}}\sim
\left(\frac{P}{c}\right)^{\,\beta}
\end{equation}
where the exponent is of order $\beta \approx 0.57\;$. Therefore, even though diseconomies associated with polycentric cities scale superlinearly with population, it would be even worse if we did not let cities evolve from the monocentric case. The same reasoning applies to the consumption of gasoline and the $CO_2$ emissions. This suggests that, everything else being equal, polycentricity should be favoured for quality of life and environmental reasons.
\begin{figure}
\includegraphics[width=\linewidth]{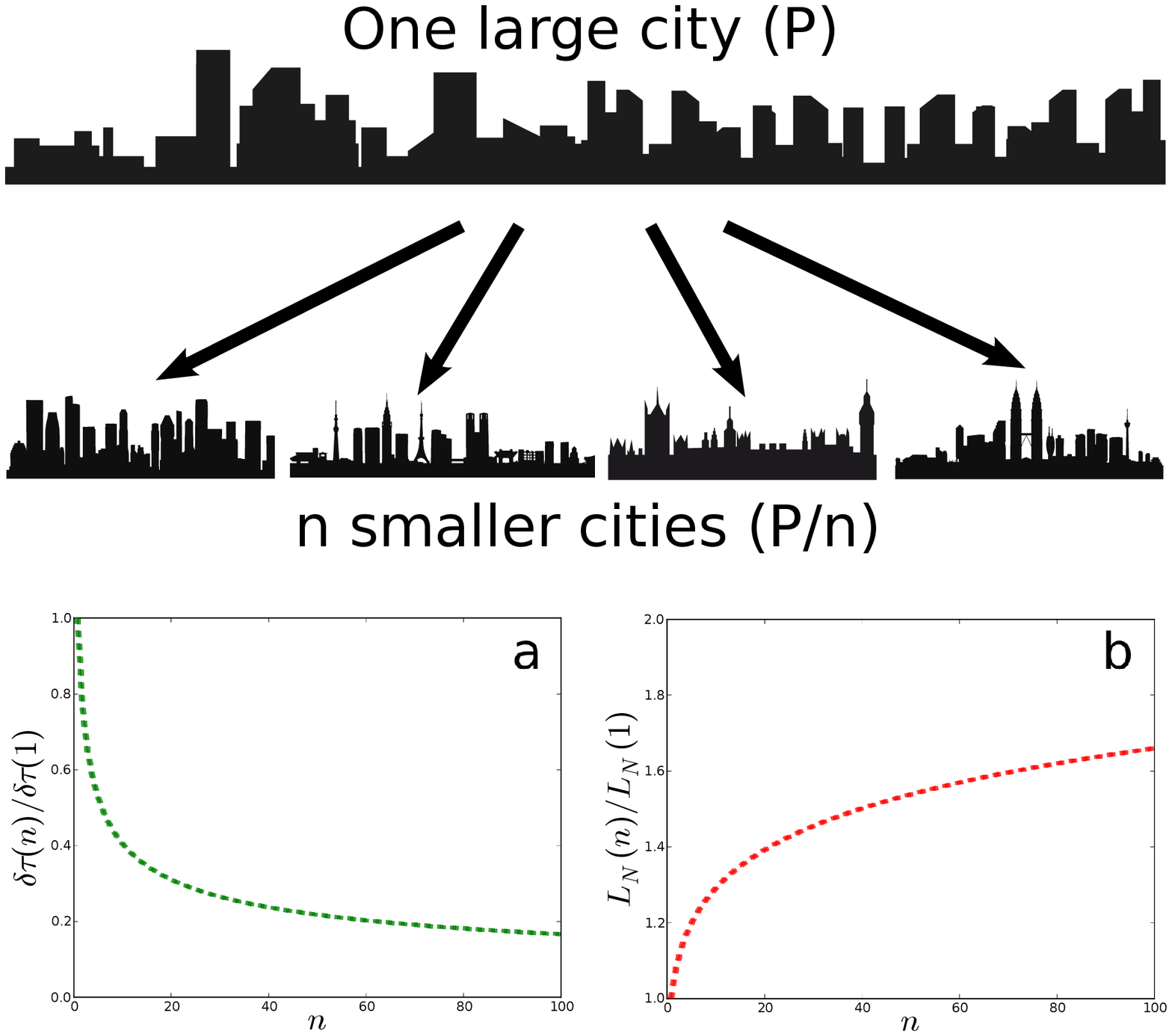}
\caption{Scaling down. We consider a population $P$ and see how indicators change when we compare it with a system with many cities and the same total population. (a) Variation of the yearly delay per capita due to congestion with the number of cities (normalised by the value $\delta\tau(1)$ corresponding to the single city case). (b) Evolution of the infrastructure length with the number of cities (normalised by the value $L_N(1)$ corresponding to the single city case). Relative gains in terms of commuting time per person decrease faster than infrastructure costs increase, suggesting that life in cities could be improved at a relatively low cost by decentralisation.}
\end{figure}

\subsection{Megapolis versus urban villages} Also, given the superlinear behaviour of the diseconomies associated with living in cities, it is clear that we would be better off living in several smaller cities rather than a single huge city. However, due to the economies of scale realised in large cities, we can wonder whether this is also economically reasonable. If we assume that the total cost of a city of population $P$ is the sum of its infrastructure cost and the economical losses due to congestion we have
\begin{equation}
C_T(P) = \epsilon_I \; L_N(P) + \epsilon_C \; \Delta t  \; \delta \tau(P)
\end{equation}
where $\epsilon_I$ is the average cost of a kilometer of roads, $\epsilon_C$ the average hourly wage and $\Delta t$ the planning horizon in years (this expression is not exhaustive, as the costs dues to $CO_2$ emissions and gasoline consumptions are not included). The infrastructure needs maintenance, and its cost depends on the planning horizon as well and can be written $\epsilon_I= \epsilon_{B} + \Delta t\; \epsilon_{M}$ where $\epsilon_B$ is the construction cost in $\$/km$ and $\epsilon_M$ the maintenance cost in $ \$/km/year$.

We assume that the population $P$ is distributed among $n$ cities of the same size $P/n$ (see Figure 4). The total lane miles for the $n$ cities reads $L^{(n)}_N(P)=n\,L_N(P/n)$ where $L_N(P)\sim\ell \sqrt{P}\left(\frac{P}{c}\right)^\delta$ is the total lane for one city.  The total congestion delay for $n$ cities is $\delta \tau_n = n\delta\tau(P/n)$ and we thus obtain the total cost $C_T(P,n)$ for $n$ cities
\begin{equation}
C_T(P,n) = n^{\, -\delta} \left[ \ell\; \epsilon_I \sqrt{\frac{n}{P}} + \tau\;\epsilon_C \Delta t \right] 
\end{equation}
The number of cities $n_{min}$ which minimises the total cost is obtained when $\frac{\mathrm{d} C}{\mathrm{d} n} = 0$, leading to (for $\Delta t\gg 1$)
\begin{equation}
n_{min} = P \left[ \left( \frac{2\delta}{1+\delta} \right) \frac{\epsilon_C}{\epsilon_M} \frac{\tau}{\ell} \right]^2
\end{equation}
(the actual number of cities is of course an integer, and can be taken as the nearest integer from $n_{min}$ for instance).  It is then economically advantageous to divide the population in several cities if $n_{min} > 2$. To illustrate this point, we compute the number of cities which would minimise the cost for a world population $P \approx 10^9$. The World Bank estimates the maintenance cost of roads to be of the order of $\epsilon_M \approx 10^5\; \$/km/year$, and the average hourly wage to be of the order of $\epsilon_C \approx 10\; \$/h$, the value of $\delta$ is taken from the measures on US data, $\delta \approx 0.27$, and $\tau / \ell \approx 10\; km/h$. We then obtain
\begin{equation}
n_{min} \approx 180
\end{equation}
which gives an average city size of $P/n\approx 5,500,000$. This result is to put in perspective with the fact that the world hosts
$40$ or so cities with over $5,500,000$ inhabitants and that this number is still increasing. 

\subsection{The most economical population distribution}

The previous results assume that we split a large city into many cities of the same size. The cities are however organized in various sizes distributed according to something that can be approximated by a Pareto distribution, as known since Zipf's work~\cite{Zipf:1949}. It is still unclear why we observe such a convergence~\cite{Batty:2006,Cristelli:2012}. We propose here a new perspective to this debate by asking: Assuming cities are distributed according to a Pareto distribution, what value of the exponent minimises the overall cost? Indeed from above the total cost for a population size $x$ is given at large times by
\begin{equation}
C_T(x) = \epsilon_M\Delta t \; L_N(x) + \epsilon_C \; \Delta t  \; \delta \tau(x)
\end{equation}
We assume that the population is distributed according to 
\begin{equation}
\mathcal{P}(x)=(\gamma-1) x^{-\gamma}\;\;\;\text{for}\;\;x\in[1,\Lambda ]
\end{equation}
with $\gamma>1$ and a cut-off population $\Lambda \gg 1$ (which is at most equal to the world's population). The average cost is then given by
\begin{equation}
\overline{C_T} = \int_1^\Lambda \mathcal{P}(x)\,C_T(x)\,dx
\end{equation}
leading to
\begin{equation}
\overline{C_T} = \frac{\Delta t\, \ell}{c^\delta}\,\left(\gamma - 1 \right) \left[ \epsilon_M\frac{\Lambda^{-\gamma+\delta+\frac{3}{2}}-1}{-\gamma+\delta+\frac{3}{2}} + \frac{\epsilon_C}{v}\frac{\Lambda^{-\gamma+\delta+2}-1}{-\gamma+\delta+2} \right]
\end{equation}
The only consistent solution is obtained for $\gamma < \delta + 2$. The dominant term for $\Lambda \gg 1$ is given by
\begin{equation}
\overline{C_T}\simeq \frac{\Delta t\,\epsilon_C\, \ell}{c^\delta}\,\left(\gamma - 1 \right) \left[ \frac{\Lambda^{-\gamma+\delta+2}}{-\gamma+\delta+2} \right]
\end{equation}
The optimal power law distribution minimizes the average cost and is such that $\frac{{\rm d} \overline{C_T}}{{\rm d} \gamma} = 0$. We obtain the following equation
\begin{equation}
\frac{1+\delta}{\delta+2-\gamma} = \left(\gamma - 1\right) \ln \Lambda
\end{equation}
and in the limit $\Lambda \gg 1$ we obtain the optimal value for $\gamma$
\begin{equation}
\gamma^* = 2+\delta - \frac{1}{\ln \Lambda}
\end{equation}
Numerically, $\delta \approx 0.32$ and $\Lambda \approx 10^{9}$, leading to $\gamma^* \approx 2.27$. It is interesting to note that this value would lead to a rank-plot exponent ($\approx 0.78$) not far from those measured on different countries around the world~\cite{Soo:2005}. Although we do not pretend that the above reasoning provides a definitive answer to the Zipf puzzle, it nevertheless suggests that the broad diversity of population might derive from economical considerations, and that there may be a connection between the Zipf law exponent and optimality considerations.

\subsection{Outlook}

The superlinear increase of congestion delay with population, and thereby of gasoline consumption and of $CO_2$ emissions, has terrible consequences on the economy, the environment, health and well-being. The outlook is nothing short of grim in our ever-urbanising world. As the proportion of human beings living in cities dramatically increases --the UN expects the world population to be $67\%$ urban in 2050~\cite{UN:2011}-- wages are likely to increase~\cite{Bettencourt:2007} but not enough to compensate for the negative effects of congestion. As a result, if the individual car stays the dominant transportation mode, cities will put more strain on people's life, while acting as catalysts for the production of $CO_2$ greenhouse gas, responsible for an overall increase of the planet's temperature~\cite{Oreskes:2004}. It is currently believed that advantages associated with living in a large city outweigh the costs. Our results reveal however the existence of very rapidly growing problems such as congestion and $CO_2$ emissions, which inevitably begs the question of the sustainability of large cities. It might be time to cut down considerably the use of individual vehicles, or to consider the possibility of living in smaller or medium sized cities: the infrastructure costs ($L_N$) may be larger, but the impact on the environment ($CO_2$ emissions) and on the well-being of people (delays in congestion) would be beneficial (see Figure 3).

The most striking fact about the above results is that despite the apparence of complexity that is conveyed by cities, most of their structure can be explained by the very simple and universal desire for the best achievable balance between income and commuting costs. Our model unifies mobility patterns, spatial structure of cities and allometric scalings in a framework that can be built upon. More work is needed in order to integrate information about firm locations, the influence of public transportation on mobility patterns~\cite{Roth:2011}, the effect of the integration of cities into urban systems~\cite{Rozenblat:2007}, to understand the fluctuations around the average trends, and to test the validity of the model on different sets of data. We believe however that the results presented here represent a crucial step towards a scientific understanding of cities.

%% Optional Materials and Methods Section
%% The Materials and Methods section header will be added automatically.
%% Enter any subheads and the Materials and Methods text below.
%%%%%%%%%%%%%%%%%%%%%%%%%
\section{Methods}

\subsection{Data} As recently stressed in~\cite{Arcaute:2013}, when trying to identify patterns accross cities, one must be careful and consistent in the definition of city boundaries. These authors indeed found out that the scaling exponents measures for several quantities are usually sensitive to the definition chosen for the city. In order to make our results reproducible, we detail in the following the data source and the corresponding city definition.\\

{\it Total distance driven and lane miles} The daily commuting vehicle-miles as well as the total lane miles data were obtained for the year 2011 from the Federal Highway Administration for 441 Urban Areas (as defined by the Census Bureau) in the US.\\

{\it Area} The surface area data were obtained for the year 2010 from the Census Bureau for 3540 Urban Areas (as defined by the Census Bureau) in the US. It is interesting to note that the dependence of the surface area of Metropolitan Statistical Areas with population is a lot less clear-cut, implying that, with respect to surface area, the definition of UAs delineates more coherent systems than the definition of MSAs.\\

{\it Values of $\ell$} In order to compute a value for $\ell = \frac{s}{t}$, we use for $s$ the average wage at the county level, provided by the Bureau of Labor Statistics. For $t$, the transportation cost per unit distance, we use the average gas price per state as given by the U.S. Energy Information Administration, and assume that all vehicles burn the same quantity of gas per unit distance on average. Interestingly, while we have assumed a constant $\ell$ throughout this paper, we have noticed that its effect on the different scalings was not negligible (Compare the results for $L_N$ and $A$ between Table 1 and Table 2 for instance), implying that $\ell$ has a small, yet non-zero dependence on the population. This probably comes from the dependence of the average wage on population~\cite{Bettencourt:2007}. We leave the investigation of this dependence for further studies.\\

{\it Total delay and $CO_2$ emissions} The excess $CO_2$ and the total delay due to traffic congestion were obtained for the year 2012 from the Urban Mobility Report for 97 Urban Areas in the US. Also, the quantity of $CO_2$ emissions due to transportation was obtained from the OECD for 268 metropolitan areas accross 28 countries for the year 2008. It is worth noting here that the US definition of Urban Area and the OECD definition of Metropolitan Area are qualitatively different, added to the fact that OECD data cover many different countries. Yet, the measured values of the exponent are compatible with each other.

As far as the United States are concerned, we present results for Urban Areas only. Indeed, when data were available for both MSA and Urban Area, we found out that the MSA data did not exhibit as clear-cut regularities as the Urban Area data did. We believe that this effect is due to the lack of a unique, quantitative definition of a city which makes  In this work, we assumed that Urban Areas designate areas which are coherent with respect to the quantities we are measuring, and leave the crucial issue of city definition for further studies.

\section{Acknowledgments} We thank Giulia Carra, Riccardo Gallotti and Thomas Louail for stimulating discussions. 
MB acknowledges funding from the EU commission through project EUNOIA (FP7-DG.Connect-318367).

\section{Author Contributions Statement}

RL, MB designed, performed research and wrote the paper.

\section{Competing financial interests}
The author(s) declare no competing financial interests.

% \section{Supplementary Information}

% As suggested by Newman and Kenworthy's now famous plot~\cite{Newman:1989}, the quantity of gasoline $Q_{gas}$ consummed in a given city should be a simple function of this city's population density

% \begin{equation}
% Q_{gas} = f(\rho)
% \end{equation} 

% where $f$ is an unknown function. While we don't have direct access to gasoline consumption data, it seems reasonable to assume that the quantity of $CO_2$ that is emitted due to transports will be proportional to the gasoline consumption. We thus plot the quantity of emitted $CO_2$ for the OECD cities as a function a city density on Fig.~\ref{fig:NK}. A fit assuming a power-law dependence gives a poor $R^2$ value compared to that we obtained for 

\bibliographystyle{prsty}

\begin{thebibliography}{99}

\bibitem{Fujita:1999} Fujita, M., Krugman, P. R., \& Venables, A. J. {\it The spatial economy: cities, regions and international trade} (The MIT press, Cambridge, USA, 1999).

\bibitem{Batty:2007} Batty, M.  {\it Cities and complexity: understanding cities with cellular automata, agent-based models, and fractals} (The MIT press, Cambridge, USA 2007)

\bibitem{Marshall:2004} Marshall, S. {\it Streets and patterns} (Routledge, Abingdon, 2004)

\bibitem{Newman:1989} Newman, P. W. \& Kenworthy, J. R. Gasoline consumption and cities: a comparison of US cities with a global survey. {\it Journal of the American Planning Association} {\bf 55}, 24–37 (1989).

\bibitem{Makse:1995} Makse, H.A., Havlin, S., Stanley, H.E. Modelling urban growth. {\it Nature} {\bf  377}, 608-612 (1995).

\bibitem{Pumain:2006} Pumain, D., Paulus, F., Vacchiani-Marcuzzo, C., \& Lobo, J. An evolutionary theory for interpreting urban scaling laws. {\it Cybergeo: European Journal of Geography}. (2006)

\bibitem{Bettencourt:2007} Bettencourt, L.M.A., Lobo, J., Helbing, D., K\"uhnert, C. \& West, G.B. Growth, innovation, scaling, and the pace of life in cities. {\it Proc. Natl. Acad. Sci. U.S.A.} {\bf 104}, 7301-7306 (2007).

\bibitem{Samaniego:2008} Samaniego, H., \& Moses, M.E.,  Cities as organisms: Allometric scaling of urban road networks {\it Journal of Transport and Land Use} {\bf 1}, 21-39 (2008).

\bibitem{Rozenfeld:2008} Rozenfeld, H. D. et al. Laws of population growth. {\it Proc. Natl Acad. Sci. USA} {\bf 105}, 18702–18707 (2008).

\bibitem{Pan:2013} Pan, W., Ghoshal, G., Krumme, C., Cebrian, M., \& Pentland, A. Urban characteristics attributable to density-driven tie formation. {\it Nature communications}, {\bf 4}:1961 (2013).

\bibitem{Fujita:1982} Fujita, M. \& Ogawa, H. Multiple equilibria and structural transition of non-monocentric urban configurations. {\it Regional science and urban economics} {\bf 12}, 161-196 (1982).

\bibitem{Batty:2008} Batty, M. The Size, Scale, and Shape of Cities. {\it Science} {\bf 319}, 769–771 (2008).

\bibitem{Frasco:2013} Frasco, G.F., Sun, J., Rozenfeld, H.D. \& ben-Avraham, D. Spatially distributed social complex networks {\it ArXiv:1306.0257} (2013)

\bibitem{Bettencourt:2010} Bettencourt, L. \& West, G. A unified theory of urban living. {\it Nature} {\bf 467}, 912–913 (2010).

\bibitem{Bettencourt:2013} Bettencourt, L. M. A. The Origins of Scaling in Cities. {\it Science} {\bf 340}, 1438–1441 (2013).

\bibitem{Gonzalez:2008} Gonzalez, M., Hidalgo, C.A., Barabasi, A.-L.. Understanding individual human mobility patterns. {\it Nature} {\bf 453}, 779-782 (2008).

\bibitem{Louf:2013} Louf, R. \& Barthelemy, M. Modeling the polycentric transition of cities. {\it Physical Review Letters} {\bf 111}, 198702 (2013)

\bibitem{Strano:2012} Strano, E., Nicosia, V., Latora, V., Porta, S. \& Barthelemy, M. Elementary processes governing the evolution of road networks {\it Scientific Reports} {\bf 2}:296 (2012).

\bibitem{Barthelemy:2013} Barthelemy, M., Bordin, P., Berestycki, H., \& Gribaudi, M. Self-organization 
versus top-down planning in the evolution of a city. {\it Scientific Reports} {\bf  3}:2153 (2013).

\bibitem{Wilkerson:2013} Wilkerson, G., Khalili, R. \& Schmid, S. Urban Mobility Scaling: Lessons from Little Data. {\it arXiv:1401.0207} (2013). 

\bibitem{Glaeser:2001} Glaeser, E. L. \& David C. M. Cities and skills. {\it Journal of Labor Economics} {\bf 19}, 316–342 (2001).

\bibitem{Dyson:1962} F.J. Dyson, {\it Journal of Mathematical Physics}
  {\bf 3}, 140-156 (1962).

\bibitem{McMillen:2003} McMillen, D. P. \& Smith, S. C. The number of subcenters in large urban areas. {\it Journal of Urban Economics} {\bf 53}, 321–338 (2003).

\bibitem{Bouchaud:2008} Bouchaud, J.-P. Economics needs a scientific revolution. {\it Nature} {\bf 455}, 1181-1181 (2008).

% \bibitem{Barenblatt} Barenblatt, G. I. {\it Scaling, self-similarity, and intermediate asymptotics: dimensional analysis and intermediate asymptotics.} Cambridge University Press. Cambridge. (1996)

\bibitem{Barthelemy:2011} Barthelemy, M. Spatial networks. {\it Physics Reports} {\bf 499}, 1–101 (2011).

%\bibitem{Newman:1999} Newman, P., \& Kenworthy, J. {\it Sustainability and cities: overcoming automobile dependence.} %Island Press. Washington, DC. (1999)

\bibitem{Oliveira:2014} Oliveira, E. A., Andrade Jr., J. S. \& Makse, H. A. Large cities are less green. {\it Scientific Reports} {\bf 4} 4235 (2014)

\bibitem{Fragkias:2013} Fragkias, M., Lobo, J., Strumsky, D. \& Seto, K. C. Does Size Matter? Scaling of CO2 Emissions and U.S. Urban Areas. {\it PLoS ONE} {\bf 8}, e64727 (2013).


\bibitem{Rybski} Rybski, D., Sterzel, T., Reusser, D.E., Winz, A.-L., Fichtner, C., Kropp, J.P. Cities as nuclei of sustainability? {\it  arXiv:1304.4406} (2013).

\bibitem{Zipf:1949} Zipf, G.K. {\it Human behavior and the principle of least effort} (Addison-Wesley Press, Cambridge, 1949).

\bibitem{Batty:2006} Batty, M. Rank clocks. {\it Nature} {\bf 444}, 592–596 (2006).

\bibitem{Cristelli:2012} Cristelli, M., Batty, M. \& Pietronero, L. There is More than a Power Law in Zipf. {\it Scientific Reports} {\bf 2}, 812 (2012).

\bibitem{Soo:2005} Soo, K. T. Zipf’s Law for cities: a cross-country investigation. {\it Regional science and urban Economics} {\bf 35}, 239–263 (2005).

\bibitem{UN:2011} United Nations, The world urban propects: The 2011 revision. (2011)

\bibitem{Oreskes:2004} Oreskes, N. The scientific consensus on climate change. {\it Science} {\bf 306}, 1686 (2004)

\bibitem{Roth:2011} Roth, C., Kang, S. M., Batty, M. \& Barthelemy, M. Structure of Urban Movements: Polycentric Activity and Entangled Hierarchical Flows. {\it PLoS ONE} {\bf 6}, e15923 (2011).

\bibitem{Rozenblat:2007} Rozenblat, C., \& Pumain, D. ``Firm linkages, innovation and the evolution of urban systems'' in {\it Cities in globalization: practices, policies and theories}, [130-156]  (Routledge, London, 2007).

\bibitem{Arcaute:2013} Arcaute, E. et al. City boundaries and the universality of scaling laws. {\it arXiv:1301.1674} (2013)

\bibitem{DataUSA1} Census bureau: https://www.census.gov/ (date of access:01/04/2014)

\bibitem{DataUSA2} Urban mobility report: http://mobility.tamu.edu/ums/ (date of access: 01/04/2014)

\bibitem{DataUSA3} Federal Highway administration: https://www.fhwa.dot.gov/ (date of access: 01/04/2014).

%Data about the area and population of urbanised areas can be found  on the Census Bureau website. Data about congestions in Metropolitan Statistical Areas %can be found in the Urban Mobility Report. Data about the total lane miles and the daily total miles driven in urbanized areas can be found on on the %Federal Highway administration website.

\bibitem{OECD} http://measuringurban.oecd.org (Date of access: 01/04/2014).

\end{thebibliography}

\clearpage

\section{Tables}

\begin{table*}
\begin{tabular}{|c|c|c|l|}
\hline
Quantity & Naive exponent &  Measured value\\
\hline
$A $ & $1$ & $0.85 \pm 0.011\; (r^2=0.93)$\\
\hline
$L_N / \sqrt{A}$ & $0.5$ & $0.42 \pm 0.02\; (r^2=0.83)$\\
$L_N $ & $1$ & $0.86 \pm 0.02\;(r^2=0.92)$\\ 
\hline
$L_{tot} / \sqrt{A}$ &  $[1/2,1]$ & $0.60 \pm 0.03\; (r^2=0.90)$\\
$L_{tot} /P$ &  $1$ & $0.03 \pm 0.02\; (r^2=0.04)$\\
\hline
\end{tabular}
\caption{This table displays the value of the exponent governing the behavior with the population $P$ obtained by naive arguments and the value obtained from empirical data. The discrepancies reveal the failure of the naive scaling arguments and the necessity to go further and model mobility patterns. The data used for this table can be found in \cite{DataUSA1,DataUSA2,DataUSA3}.}
\label{table:naive}
\end{table*}

\begin{table*}
\begin{tabular}{|c|c|c|c|c|}
\hline
Quantity & Theoretical dependence on $P$ & \multicolumn{2}{|c|}{Predicted value of the exponents} & Measured value\\
  $\;\;$  & in self-consistant case & self-consistant case & fitting case & \\ \hline
$L_{tot}$ & $P$ & $1$ & $1$ & $1.03 \pm 0.03\;(r^2=0.95)$\\ 
$A / \ell^2$ & $\left( \frac{P}{c} \right)^{\,2\,\delta}$ & $2 \delta = 0.64$ & $a = 0.85$ & $0.853 \pm 0.011\; (r^2=0.93)$~\cite{DataUSA1,DataUSA2,DataUSA3}\\
$L_N / \ell$ & $\sqrt{P}\; \left(\frac{P}{c}\right)^{\,\delta}$ & $\frac{1}{2} + \delta = 0.82$ & $\frac{1+a}{2} = 0.93$ & $0.765 \pm 0.033\; (r^2=0.92)$~\cite{DataUSA1,DataUSA2,DataUSA3}\\
$\delta \tau / \tau$ & $P\; \left( \frac{P}{c} \right)^{\, \delta}$ & $1 + \delta = 1.32$ & $1.22$ & $1.270 \pm 0.067\; (r^2=0.97)$~\cite{DataUSA1,DataUSA2,DataUSA3}\\
$Q_{gas,CO_2}/\ell$  & $P\left( \frac{P}{c}\right)^{\delta}$ & $1+\delta=1.32$ & $1.22$ & $1.262 \pm 0.089\; (r^2=0.94)$~\cite{DataUSA1,DataUSA2,DataUSA3}\\
& & & & $1.212 \pm 0.098\, (r^2=0.83)$~\cite{OECD}\\
& & & & $1.33 \pm 0.03$~\cite{Oliveira:2014}\\
\hline \hline
$L_N / \sqrt{A}$ & $\sqrt{P}$ & $0.5$ & $0.5$ & $0.42 \pm 0.02\;(r^2=0.83)$~\cite{DataUSA1,DataUSA2,DataUSA3}\\
$L_{tot} / \sqrt{A}$ &  $P\;
\left(\frac{P}{c}\right)^{-\delta}$ & $1 - \delta = 0.68$ & $1-a/2=0.58$ & $0.595 \pm 0.026\; (r^2=0.90)$~\cite{DataUSA1,DataUSA2,DataUSA3}\\
\hline
\end{tabular}
\caption{This table displays the predicted theoretical behavior and
  the empirical observations versus the population size $P$ for
  different quantities: $L_{tot}$ is the daily total driven distance,
  $A$ is the area of the city, $L_{N}$ is the total length of the road
  network, $\delta\tau$ is the daily total delay due to congestion,
  $Q_{gas}$ is the yearly total consumption of gasoline and $Q_{CO_2}$
  is the total $CO_2$ emissions emitted yearly due to
  transportation. In the third column, we show the predicted values of
  the exponent of $P$ using the value of $\alpha$ measured on US
  employment data~\cite{Louf:2013}, and in the fourth column, the value of the
  exponents directly measured on data about US and OECD cities. The
  measured values are in good agreement with the prediction. In
  particular, the exponents for $L_N$ and $\delta \tau$ are consistent
  with our prediction that their difference should be $1/2$.}
\label{table:results}
\end{table*}

\end{document}